\documentclass[twocolumn,english,aps,pra,showpacs,superscriptaddress]{revtex4-1}
\usepackage[T1]{fontenc}
\usepackage[latin9]{inputenc}
\usepackage{color}
\usepackage{babel}
\usepackage{amsmath}
\usepackage{amssymb}
\usepackage{graphicx}
\usepackage{esint}
\usepackage[unicode=true,
 bookmarks=true,bookmarksnumbered=false,bookmarksopen=true,bookmarksopenlevel=1,
 breaklinks=false,pdfborder={0 0 0},backref=false,colorlinks=true]
 {hyperref}

\makeatletter
\@ifundefined{textcolor}{}
{%
 \definecolor{BLACK}{gray}{0}
 \definecolor{WHITE}{gray}{1}
 \definecolor{RED}{rgb}{1,0,0}
 \definecolor{GREEN}{rgb}{0,1,0}
 \definecolor{BLUE}{rgb}{0,0,1}
 \definecolor{CYAN}{cmyk}{1,0,0,0}
 \definecolor{MAGENTA}{cmyk}{0,1,0,0}
 \definecolor{YELLOW}{cmyk}{0,0,1,0}
}

\makeatother

\begin{document}

\title{Pump Electron-Positron Pairs from Well Potential }

\author{Qiang Wang}

\affiliation{National Laboratory of Science and Technology on Computational Physics,
Institute of Applied Physics and Computational Mathematics, Beijing
100088, China}

\author{Jie Liu}

\affiliation{National Laboratory of Science and Technology on Computational Physics,
Institute of Applied Physics and Computational Mathematics, Beijing
100088, China}

\affiliation{HEDPS, Center for Applied Physics and Technology, Peking University,
Beijing 100871, China \\
and IFSA collaborative Center of MoE College of Engineering, Peking
University, Beijing 100871, China}

\author{Li-Bin Fu}

\email{lbfu@iapcm.ac.cn}

\affiliation{National Laboratory of Science and Technology on Computational Physics,
Institute of Applied Physics and Computational Mathematics, Beijing
100088, China}

\affiliation{HEDPS, Center for Applied Physics and Technology, Peking University,
Beijing 100871, China \\
and IFSA collaborative Center of MoE College of Engineering, Peking
University, Beijing 100871, China}
\begin{abstract}
In this paper we show that electron-positron pairs can be pumped inexhaustibly
with a constant production rate from the one-dimensional well potential
with oscillating depth or width. Bound states embedded in the the
Dirac sea can be pulled out and pushed to the positive continuum,
become scattering states. Pauli block, which dominant the saturation
of pair creation in the static super-critical well potential, can
be broken by the ejection of electrons. We find that the width oscillating
mode is more efficient that the depth oscillating mode. In the adiabatic
limit, pair number as a function of upper boundary of the oscillating,
will reveal the diving of the bound states.
\end{abstract}

\pacs{03.65.Pm,12.20.-m, 02.60.-x }

\maketitle

\section{introduction}

Since Einstein's relativistic theory tell that the matter can convert
into energy, the possibility of converting energy into matter, i.e.,
the electron-positron pair, as predicted by Dirac in quantum electrodynamics
\citep{Dirac}, has attracted a great deal of interest\citep{RMP_Keitel_2012}.
In presence of a static and uniform electric field, the quantum electrodynamic
(QED) vacuum may break down and decay into electron-positron pairs
due to a quantum tunneling effect \citep{breakdown1_1931,breakdown2_1936,breakdown3_1951}.
The critical Schwinger field is $E_{c}=m^{2}c^{3}/\left(\left|e\right|\hbar\right)$,
which can accelerate the electron to an energy of the order of its
rest mass on its Compton wavelength $\lambda_{C}=\hbar/mc$, where
$m$ is the electron mass. Starting from the works of Brezin and Popov
et.al \citep{gener_Sch_time_1,gener_Sch_time_2,gener_Sch_time_3},
the Schwinger mechanism was generalized to time dependent fields \citep{gener_Sch_time_4,gener_Sch_time_5,gener_Sch_time_Keitel_review,gener_Sch_time_6,gener_Sch_time_7,gener_Sch_time_8,gener_Sch_time_9,gener_Sch_time_10}
, where another mechanism may be responsible for the pair creation.
If the frequency of the alternating field exceeds the gap $2mc^{2}$,
electrons in Dirac sea can transit to positive states and pairs are
triggered. Experimentally, pairs can be generated by the relativistic
heavy-ion collisions\citep{heavy_ion_collisions_1993} or the collision
of an intense laser pulse and a 46 Gev electron beam\citep{SLAC_1997},
but pairs created from pure laser light has not been observed until
now. 

Recently, various numerical approach were developed to deal with the
time dependent Dirac equation\citep{Quant_dyna_re_ele_Keitel_2004,2d_code_Dirac_Keitel_2008,splitoperatormethod},the
Klein paradox\citep{Klein_paradox_su_2004,Klein_paradox_NQFT_Su_2005},
the Zitterbewegung\citep{ZB_su_2004}, and the pair production process
\citep{timing_Su_2006,Dynamics_Bound_States_Su_2005,boundstate_channel_close_su_2014,suotang2013,fit_rate,rev_nQFT_Su_2010}.
The one-dimensional well potential, specially, for its simplicity,
is studied extensively\citep{Dynamics_Bound_States_Su_2005,boundstate_channel_close_su_2014,suotang2013,MJiang_2013,Degeneracies}.
The super critical well potential has bound states embedded in the
negative continuum can cause spontaneous electron-positron pair creation.
Theoretical investigations are expected to make the physics of the
creation clear and predict a higher generation rate. However, The
Pauli exclusion principle will block further creation once the bound
states are occupied, resulting a asymptotic saturation behavior\citep{Dynamics_Bound_States_Su_2005,boundstate_channel_close_su_2014,Degeneracies}.
Motivated by this requirement and a better understanding of the pair
creation process in the one-dimensional well potential, we examine
the pair creation in a well with its width or depth oscillating. By
oscillating the width and depth, the transfer channels for population
are opened and closed alternately. The electrons confined in the well
will be released and the Pauli block become invalid. This can lead
to a non-vanishing production rate, which means that pairs can be
pumped inexhaustibly form the well.

This paper is organized as follows. In Sec.\mbox{II} we present the
model and the numerical method we employed. The well potential is
set to be oscillating in two modes, the width oscillating mode and
the depth oscillating mode. The energy spectrum is plotted as a function
of the width or the depth. In Sec.\mbox{III}. we discuss the pair
production process in both two modes. The time evolution of pair number,
spacial density and pumping rate are studied. We also investigate
the adiabatic limit of the oscillating. In the last section we give
a brief summary.

\section{model and method}

\subsection{Model: one-dimensional well potential with oscillating depth or width}

In one dimension, the time evolution of the Heisenberg field operator
$\ensuremath{\hat{\Psi}\left(z,\, t\right)}$ is given by the Dirac
equation ( without spin, for simplicity in this paper ) \citep{QED_strongfield_Greiner_1985}

\begin{equation}
i\frac{\partial}{\partial t}\hat{\Psi}\left(z,\, t\right)=\left[c\boldsymbol{\sigma}_{1}\cdot\boldsymbol{\hat{p}}_{z}+c^{2}\boldsymbol{\sigma}_{3}+V\left(z,\, t\right)\right]\hat{\Psi}\left(z,\, t\right).\label{eq:Dirac_Eq}
\end{equation}
$\boldsymbol{\sigma}_{1,\,}\boldsymbol{\sigma}_{3}$ are Pauli matrices,
$c$ is the speed of light in vacuum, $V\left(z,\, t\right)$ is the
external potential. The atomic units ({[}a.u.{]}) is used in this
paper: $m=\hbar=e=1$, $c=1/\alpha\approx137.0359991$, $\alpha$
is fine-structure constant, Compton wave length of electron is $\lambda_{C}=1/c$.
The Hamiltonian of the system is $H=\left[c\boldsymbol{\sigma}_{1}\cdot\boldsymbol{\hat{p}}_{z}+c^{2}\boldsymbol{\sigma}_{3}+V\left(z,\, t\right)\right]$.
We define the potential as

\begin{equation}
V\left(z,t\right)=\frac{V_{0}\left(t\right)}{2}\left[\tanh\left(\frac{z-\frac{W\left(t\right)}{2}}{D}\right)-\tanh\left(\frac{z+\frac{W\left(t\right)}{2}}{D}\right)\right].\label{eq:Vzt}
\end{equation}
$D$ is the width of potential edge (a measure of the width of the
electric field), and we set $D=0.3\lambda_{C}$. The numerical box
size is set to $L=2.5$. 

The potential width $W\left(t\right)$ and the depth $V_{0}\left(t\right)$
( positive, but note that the potential $V\left(z,\, t\right)$ is
negative in the center and zero elsewhere) are set to two modes: (1)
the \textbf{W-oscillating mode}: $V_{0}$ is constant, $W\left(t\right)=W_{1}+\frac{1}{2}\left(W_{2}-W_{1}\right)\left[1+\sin(\omega_{W}(t)-\pi/2)\right]$
; (2) the \textbf{V-oscillating mode}\textbf{\textsl{:}} $W$ is constant,
$V_{0}\left(t\right)=V_{1}+\frac{1}{2}\left(V_{2}-V_{1}\right)\left[1+\sin(\omega_{V}(t)-\pi/2)\right]$.
In this paper we assume $W_{1}=0$ and $V_{1}=0$, then $W\left(t\right)$
( or $V_{0}\left(t\right)$) varies as a sine function between zero
and its upper boundary $W_{2}$ ( or $V_{2}$), and the turning on
process is from zero point with first order derivative equal to zero.
In the following numerical simulation, we choose the total evolution
time to be the period ( $T_{W}$ or $T_{V}$ ) of the oscillating
$W(t)$ or $V\left(t\right)$ multiples an integer, to make the potential
turning off is finished with first order derivative equal to zero.

\begin{figure}
\includegraphics[scale=0.65]{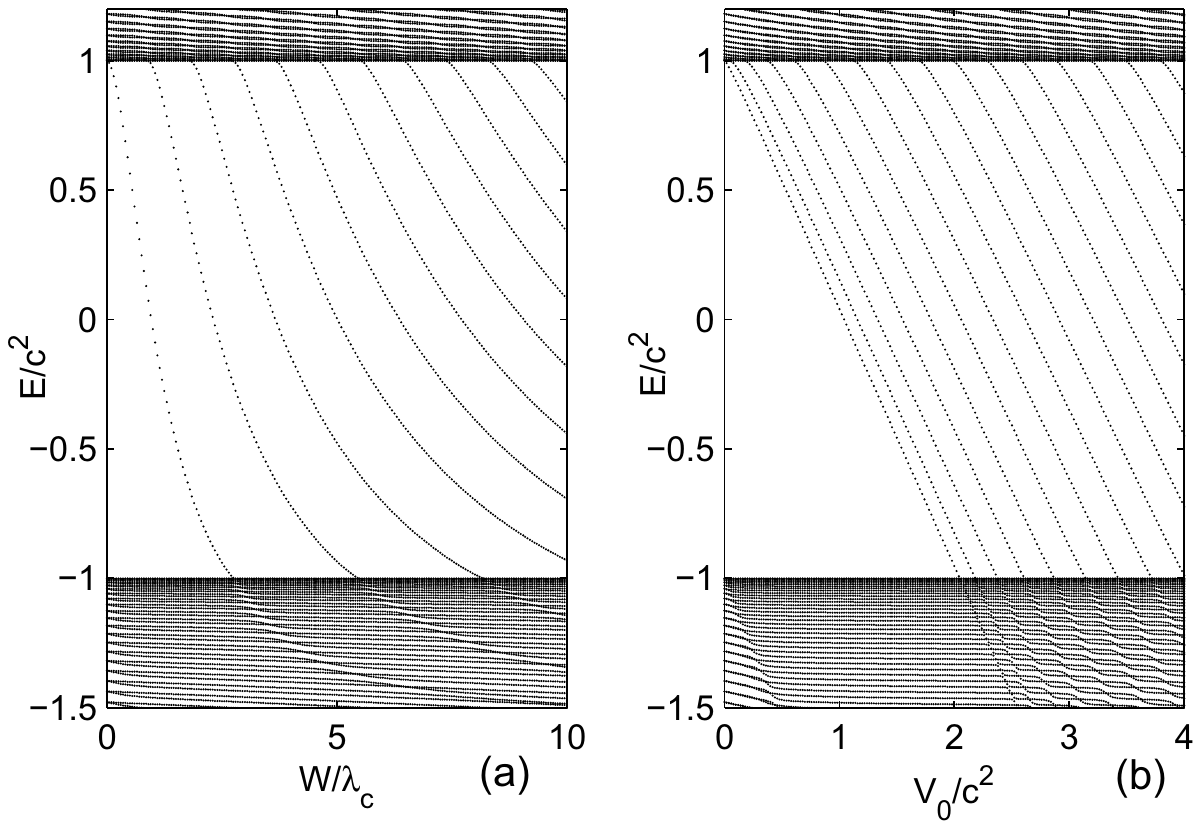}

\protect\caption{The energy spectrum of the total Hamiltonian as a function of the
width or the depth of the potential. (a), $\ensuremath{V_{0}=2.53c^{2}}$,
as $W$ increasing, the bound states dive into the Dirac sea at $W=2.79,5.51,8.21...$(in
units of $\lambda_{C}$, the electron Compton wavelength). (b), $W=10\lambda_{C}$,
as $V_{0}$ increasing, the bound states dive into the Dirac sea at
$V_{0}=2.05,2.19,2.38,2.62,2.87,3.15,3.43,3.73,...$(in units of $c^{2}$).
\label{fig:spectrum}}
\end{figure}

The numerically energy spectrum of the total Hamiltonian of finite-size
(length, for one dimension) are presented in Fig. \ref{fig:spectrum}
for varying $W$ and $V_{0}$, which can be a schematic of the real
one-dimensional system. In Fig. \ref{fig:spectrum}, we show the critical
width or depth, and the behavior of 'diving' of the bound states into
the negative continuum. For example, if $\ensuremath{V_{0}=2.53c^{2}}$,
there are bound states embedded and then pair can be spontaneously
triggered only when $W>2.79\lambda_{C}$.

\subsection{Method: the numerical quantum field theoretical approach}

In recent years, numerical quantum field theoretical approach \citep{rev_nQFT_Su_2010}
has been established to overcome the single particle picture described
by quantum mechanics and the mathematical difficulty of quantum electrodynamics.
In this section we will briefly review this method and describe that
how do we deal with the model in this paper.

The field operator can be expressed in terms of the electron annihilation
and positron creation operators as\citep{rev_nQFT_Su_2010}

\begin{eqnarray}
\hat{\Psi}\left(z,\, t\right) & = & \sum_{p}\hat{b}_{p}W_{p}\left(z,\, t\right)+\sum_{n}\hat{d}_{n}^{\dagger}W_{n}\left(z,\, t\right)\\
 & = & \sum_{p}\hat{b}_{p}\left(t\right)W_{p}\left(z\right)+\sum_{n}\hat{d}_{n}^{\dagger}\left(t\right)W_{n}\left(z\right),
\end{eqnarray}
in which $W_{p(n)}\left(z\right)=\left\langle z|p(n)\right\rangle $
is the solution of the filed-free Dirac Hamiltonian ( $V\left(z,\, t\right)=0$),
$W_{p(n)}\left(z,\, t\right)=\left\langle z|p(n)\left(t\right)\right\rangle $
is the time dependent solution of the Dirac equation \eqref{eq:Dirac_Eq},
and the term $\sum_{p(n)}$ denotes the summation over all states
with positive (negative) energy. The eigenstates of the filed-free
Hamiltonian are

\begin{eqnarray}
W_{p}\left(z\right) & = & \frac{e^{ipz}}{\sqrt{4\pi E}}\begin{bmatrix}\sqrt{E+c^{2}}\\
sign\left(p\right)\sqrt{E-c^{2}}
\end{bmatrix}\\
W_{n}\left(z\right) & = & \frac{e^{inz}}{\sqrt{-4\pi E}}\begin{bmatrix}-sign\left(n\right)\sqrt{-E-c^{2}}\\
\sqrt{-E+c^{2}}
\end{bmatrix},
\end{eqnarray}
where $E_{p}=\sqrt{c^{4}+p^{2}c^{2}}$, and $E_{n}=-\sqrt{c^{4}+n^{2}c^{2}}$
respectively. The time dependent single particle wave function $W_{p(n)}\left(z,\, t\right)$
can be got by introducing the time-evolution operator $\hat{U}\left(t_{2},t_{1}\right)=\hat{T}exp\left(-\frac{i}{\hbar}\int_{t_{1}}^{t_{2}}dt'\hat{H}\left(t'\right)\right)$, 

\begin{equation}
W_{p(n)}\left(z,\, t\right)=\hat{U}\left(t,\, t=0\right)W_{p(n)}\left(z\right),
\end{equation}
where $\hat{T}$ denotes the Dyson time ordering operator. In this
paper, we use the numerical split operator technique \citep{Quant_dyna_re_ele_Keitel_2004,2d_code_Dirac_Keitel_2008},
then

\begin{eqnarray}
W\left(t+dt\right) & \approx & e^{-iHdt}W\left(t\right)\nonumber \\
 & = & e^{-i\frac{dt}{2}H_{\partial}}e^{-idtH_{z}}e^{-i\frac{dt}{2}H_{\partial}}+O\left(dt^{3}\right),
\end{eqnarray}
with 

\begin{eqnarray}
H_{\partial} & = & c\boldsymbol{\sigma}_{1}\cdot\boldsymbol{\hat{p}}_{z}+c^{2}\boldsymbol{\sigma}_{3},\\
H_{z} & = & V\left(z,\, t\right).
\end{eqnarray}
Practically, since the derivation ( the momentum operator ) can be
implemented by replacing the operator $\boldsymbol{\hat{p}}_{z}$
with its value $k_{z}$ in momentum space, the evolution operation
has the following form 

\begin{eqnarray}
e^{-i\frac{dt}{2}H_{\partial}}W\left(t\right) & = & \mathcal{F}^{-1}\left[\cos\left(\phi\right)\right.\nonumber \\
 &  & \left.-i\sin\left(\phi\right)\frac{\sigma_{1}\cdot k_{z}+c\sigma_{3}}{\sqrt{c^{2}+k_{z}^{2}}}\right]\mathcal{F}W\left(t\right),\\
e^{-idtH_{z}}W\left(t\right) & = & \left[\cos\left(V\left(t\right)dt\right)\right.\nonumber \\
 &  & \left.-i\sin\left(V\left(t\right)dt\right)\right]W\left(t\right),
\end{eqnarray}
where $k_{z}$ is momentum, $\phi=\frac{cdt}{2}\sqrt{c^{2}+k_{z}^{2}},$
and $\mathcal{F}\left(\mathcal{F}^{-1}\right)$ is Fourier transformation
(inverse Fourier transformation).

Then, after the time dependent field operator $\ensuremath{\hat{\Psi}\left(z,\, t\right)}$
can be calculated, the number, the spacial distribution of electrons
created from the vacuum ( defined as $\hat{b}_{p}\left\Vert vac\right\rangle =0$,
$\hat{d}_{n}\left\Vert vac\right\rangle =0$) are obtained from the
positive part of the field operator, 

\begin{eqnarray}
N^{el.}\left(t\right) & = & \left\langle vac\right\Vert \hat{\Psi}^{(+)\dagger}\left(x,\, t\right)\hat{\Psi}^{(+)}\left(x,\, t\right)\left\Vert vac\right\rangle \nonumber \\
 &  & =\sum_{pn}\left|U_{pn}\left(t\right)\right|^{2},\label{eq:N_pair}\\
N_{z}^{el.}\left(t\right) & = & \sum_{n}\left|\sum_{p}U_{pn}\left(t\right)W_{p}\left(z\right)\right|^{2},\label{eq:N_z_e}
\end{eqnarray}
where $U_{pn}\left(t\right)=\left\langle W_{p}\left(z\right)|W_{n}\left(z,\, t\right)\right\rangle =\int dxW_{p}^{*}\left(z\right)W_{n}\left(z,\, t\right)$.
The pair number $N\left(t\right)$ is equal to the electron number
$N^{el.}\left(t\right)$.

The spacial distribution of the created positrons can be written as 

\begin{eqnarray}
N_{z}^{po.}\left(t\right) & = & \sum_{p}\left|\sum_{n}U_{pn}\left(t\right)W_{n}\left(z\right)\right|^{2}.\label{eq:N_z_p}
\end{eqnarray}
The total positron number $N^{po.}\left(t\right)$ is equal to the
electron number $N^{el.}\left(t\right)$.

We can also get it from the negative part of the field operator by
compute the number and spacial distribution of the holes. In this
paper we use this expression ( Eq. \eqref{eq:N_z_p}) to reduce the
computational cost, because $U_{pn}$ has been calculated in Eq.\eqref{eq:N_pair}.
Furthermore, we can neglect the larger part of the momentum ( $\sqrt{k^{2}c^{2}+c^{4}}$
is far greater than $V$ and $\omega$) in the numerical simulation,
for its contribution to the matrix element $U_{pn}\left(t\right)$
is very small. In the following, the number of spatial points is $N_{z}=2048$,
and we only take $N_{p}=1024$ discrete momentum into account. 

Based on the projection of the field operator onto the the field-free
electronic states in this method and the definition of electron and
positron in the Dirac hole theory, in this paper we will present physical
quantities for all time and focus on the moments when the field is
absent.

\section{pump electron-positron pairs from the well potential }

For a well potential of depth $V_{0}$, if $V_{0}<2c^{2}$, the positive
continuum and negative continuum can not overlap. But for a super-critical
depth, $V_{0}>2c^{2}$, the domain $c^{2}-V_{0}<E<-c^{2}$ exist,
and bound states in the well are possible (which we call ' bound states
embedded in the negative continuum ' ) : their wave function do not
decrease exponentially out the well, but join a continuum wave of
the same energy $E<-c^{2}$ out the well, hence the wave function
has a non-zero probability outside. An empty bound state will spontaneously
be occupied by an electron (two, if the spin is considered) from the
filled Dirac sea, and the hole (identified as positron) will travel
away from the well to infinity \citep{QED_strongfield_Greiner_1985}.
This is the picture of spontaneous creation of electron-positron pair.
For a static well potential, electrons will fill the  embedded bound
states, and the Pauli principle will prevent further pair creation.
The number of pair created should be the number of bound states which
meet these conditions. 

For a time dependent potential, the situation is more complicated.
In paper \citep{boundstate_channel_close_su_2014}, the effect of
open and close a pair-creation channel was studied. The well depth
is fixed at $V_{0}=2.53c^{2}$, while the width $W$ varies between
$W_{1}=4.55\lambda_{C}$ and $W_{2}=6.15\lambda_{C}$. For $W=4.55\lambda_{C}$,
there is one bound state embedded in the Dirac sea, and there are
two for $W=6.15\lambda_{C}$. After enough time for saturation, the
pair number will increase as one more channel is opened, but do not
decrease as one of the two channels is closed. The reason is that
the annihilation of the pair need the electron and positron to be
in the same place, which is not satisfied because the electron remains
in the well while the positrons have left the creation zone and escaped
to the opposite direction. 

Naturally, one can propose that if the channel is opened and closed
periodically, can this mechanism will lead to a continuously pair
creation? Moreover, for fixed $W$ and varying $V_{0}$, since the
diving behavior is similar (Fig. \ref{fig:spectrum}), will something
similar happen? Motivated by these questions, we construct two oscillating
modes as described in Sec. \mbox{II}. A. Results and discussion are
as follows. 

\begin{figure}
\includegraphics[scale=0.62]{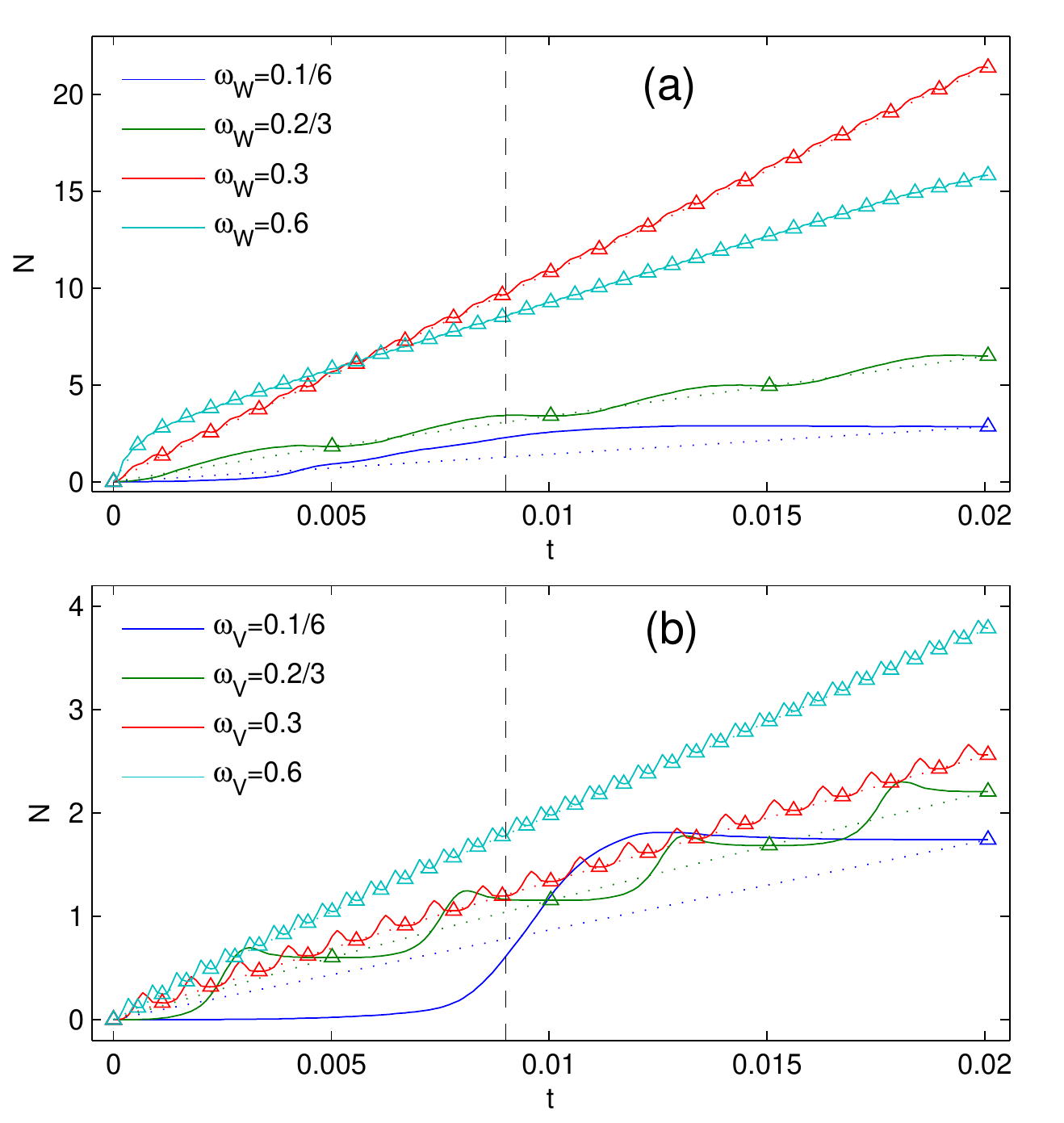}

\protect\caption{The time evolution of the total number of pairs for both W-oscillating
and V-oscillating mode.  (a), W-oscillating mode, $W_{2}=10\lambda_{C}$,$V_{0}=2.53c^{2}$;
(b), V-oscillating mode,\textbf{ $V_{2}=2.53c^{2}$}, $W=10\lambda_{C}$.
The frequency $\omega_{W}$ and $\omega_{V}$ are in units of $c^{2}$.
The dash line represent the time $t=0.009$ when the positrons arrive
the boundary, $z=\pm L/2=\pm1.25$. The triangles denote pair number
when the field is absent. The dot line just link these triangles.\label{fig:N_t}}
\end{figure}

\subsection{time evolution of pair number}

Using the method presented in Sec. \mbox{II}. B, we graph the time
evolution of the pair number defined as Eq.(\ref{eq:N_pair}) for
both W-oscillating and V-oscillating mode in Fig. \ref{fig:N_t}.
The width frequency $\omega_{W}$ and depth frequency $\omega_{W}$
are in units of $c^{2}$, and their values are assumed to be relative
low, comparing to the gap $2c^{2}$, so that the photon absorption
mechanism is not valid. The total time is $120\pi/c^{2}\approx0.02$
and the period is $T_{W}=2\pi/\omega_{W}$ or $T_{V}=2\pi/\omega_{V}$.
The dot line represent the time $t\approx L/\left(2c\right)\approx0.009$
when the particles arrive the boundary, $z=\pm L/2=\pm1.25$. Since
$W_{1}=0$ and $V_{1}=0$, if the time is an integer multiples of
the period ( $T_{W}$ or $T_{V}$ ), the system Hamiltonian degenerate
to a field free one. The triangles in Fig. \ref{fig:N_t} denote the
pair number when the field is absent. 

\textbf{W-oscillating mode: }In Fig. \ref{fig:N_t} (a), we illuminate
the total number of pairs as a function of time for $\omega_{W}=0.1/6c^{2}$,
$0.2/3c^{2}$, $0.3c^{2}$, $0.6c^{2}$. The depth $V_{0}$ is fixed
at $V_{0}=2.53c^{2}$. The width $W$ varies between $W_{1}=0$ and
$W_{2}=10\lambda_{C}$, corresponding zero and three bound states
embedded. When $W=W_{2}$, there are also eight bound states exist
in the gap, which can be associate with the pair creation\citep{Dynamics_Bound_States_Su_2005}. 

\begin{figure}
\includegraphics[scale=0.64]{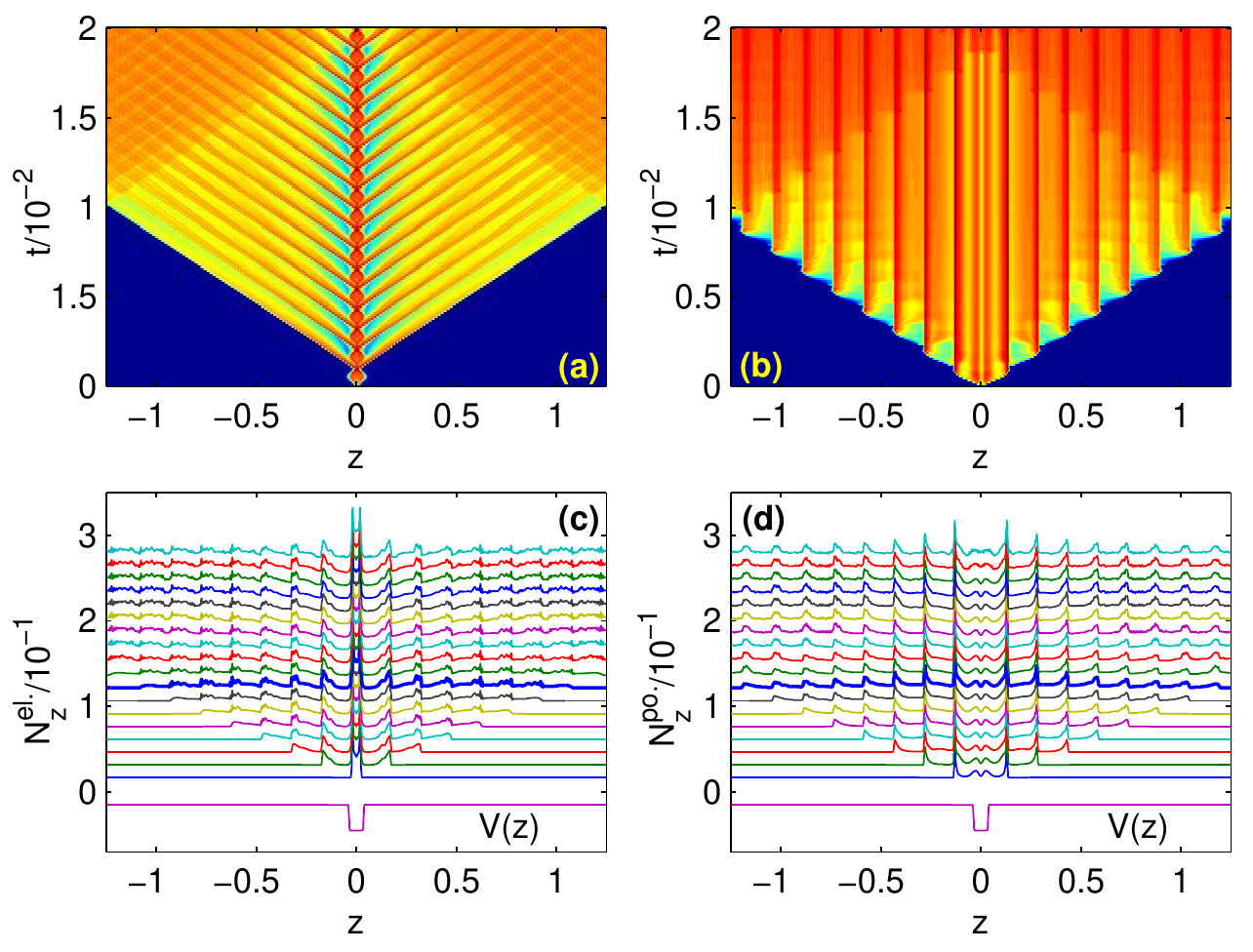}

\protect\caption{For W-oscillating mode, $\omega_{W}=0.3c^{2}$, the three dimensional
diagrams for entire time and the waterfall figures for field free
moments (the time indicated by triangles in Fig. \ref{fig:N_t} (a)),
for electron spacial density (a, c) and positron spacial density (b,
d). The thicker curve in sub-figure (c, d) mark the last cycle before
the positron arrive the boundary. The well potential $V(z)$ with
$V_{0}=2.53c^{2}$, $W=10\lambda_{C}$, are included on the bottom
for comparison. All other parameters are the same as Fig. \ref{fig:N_t}
(a).\label{fig:imagesc_W}}
\end{figure}

When $\omega_{W}=0.1/6c^{2}$, the width $W$ can only finish one
cycle in the total time $120\pi/c^{2}$. $N$ begin to arise before
$t=3.57\times10^{-3}$ , corresponding $W\left(t\right)=2.79\lambda_{C}$,
when the first bound state dive into the negative continuum. The reason
is the non-adiabatic varying width, and $N$ will begin to arise precisely
at the time when $W\left(t\right)=2.79\lambda_{C}$ in the adiabatic
case ($\omega_{W}\rightarrow0$ , see the discussion below). $N$
increases as more bound states dive in, and reach its maximum $N=2.89$
at $t=1.37\times10^{-2}$ , between $t=1.28\times10^{-2}$ and $1.47\times10^{-2}$,
at which time the third and the second bound state were pulled out
the Dirac sea. Undergoing the particle-antiparticle annihilation,
$N$ decreases but remains an appreciable value $N=2.85$ at the end.
In the latter half of this cycle, the embedded bound states depart
from the Dirac sea, return to the positive continuum, and become scattering
states. The released positrons are reflected by the numerical box
boundary, come back to the interaction region and will affect the
pair generation after. Though the effect is weak when $\omega_{W}=0.1/6c^{2}$,
it is non-ignorable when, i.e., $\omega_{W}=0.3c^{2}$ (see Fig. \ref{fig:imagesc_W}
for details). 

For $\omega_{W}=0.2/3c^{2}$ and $\omega_{W}=0.3c^{2}$ , $W$ can
finish four and eighteen cycles in the total time and the pair number
are $N=6.49$, $N=21.4$ at the end. For $t<0.009$, $W$ can finish
one and eight cycles, respectively. In each cycle, the positrons are
repulsed by the electric field to the infinity once they were generated,
while the electrons are limited in the well when the field is strong
enough and extruded out as the well is turning off, avoiding the inevitable
Pauli block in the non-varied static well construction. The non-synchronous
ejection prevent the annihilation and lead to a high production rate. 

The next cycle starts from field free and is independent on the previous
cycle. In Fig. \ref{fig:N_t}, the dot line link the triangles which
denote the pair number when the field is absent. We can find that
the pair generation denoted by the dot line is linearly depend on
time for low frequency $\omega_{W}$, for $t<0.009$. If the system
length $L$ is infinite and there is no reflection at the boundary,
the pairs can be pumped inexhaustibly with a constant production rate
from the well. Even for $\omega_{W}=0.6c^{2}$, there is nonlinear
effect at the beginning, the generation rate become stable soon. 

Due to the finite period $T_{W}$ and the bound states in the gap,
particle generation and ejection process is not monotonic with the
increase of the frequency $\omega_{W}$, see Fig. \ref{fig:N_t} (a),
However, ignoring the reflection, if the W-oscillating frequency $\omega_{W}$
is very small, we can expect a linear dependent of final pair number
on the frequency. 

\textbf{V-oscillating mode:} The number of pairs $N$ as a function
of time are presented in Fig. \ref{fig:N_t} (b), for $\omega_{V}=0.1/6c^{2}$,
$0.2/3c^{2}$, $0.3c^{2}$ and $0.6c^{2}$. The width $W$ is fixed
at $W=10\lambda_{C}$, while the depth varies between $V_{1}=0$ and
$V_{2}=2.53c^{2}$, corresponding zero and three bound states embedded.
There are also eight bound states exist in the gap when $V_{0}=V_{2}$. 

For $\omega_{W}=0.1/6c^{2}$, the first bound state dive in at $t=7.20\times10^{-3}$,
at which time there are already $N=8.83\times10^{-2}$ pair generated.
The first bound state depart the negative continuum after the second
and the third one, at $t=1.29\times10^{-2}$, when $N$ reach its
maximum $N=1.81$. Finally, there are $N=1.74$ pairs survived at
$t=120\pi/c^{2}$. For $\omega_{V}=0.2/3c^{2}$, $0.3c^{2}$, $0.6c^{2}$,
the pair number at the end are $N=2.21,$ $2.56$, $3.78$. 

Instead of pulling and pushing the walls of the well in W-oscillating
mode, in this mode it is the rising and falling bottom of the well
that control the bound states diving in and departing from the negative
continuum. It is also the non-synchronous ejection of the positrons
and electrons which dominant the pumping process. 

The dot line here indicate a linear relation between the pair number
and time. The final number is not monotonic depending on the frequency
$\omega_{V}$, and we can also expect a linear dependent of final
pair number on $\omega_{V}$ when $\omega_{V}$ is very small. 

Note that although the two modes has the same beginning and ending
parameters, the generation rate in the W-oscillating mode is much
higher.

\subsection{time evolution of spacial density }

In last section we discussed the total pair number as a function of
time in different oscillating frequency for both modes. To show the
pumping process explicitly, we compute the time evolution of spacial
density of electrons and positrons ( Eq. \ref{eq:N_z_e} and Eq.\ref{eq:N_z_p})
for $\omega_{W}=0.3c^{2}$ and $\omega_{V}=0.3c^{2}$ respectively. 

\begin{figure}
\includegraphics[scale=0.64]{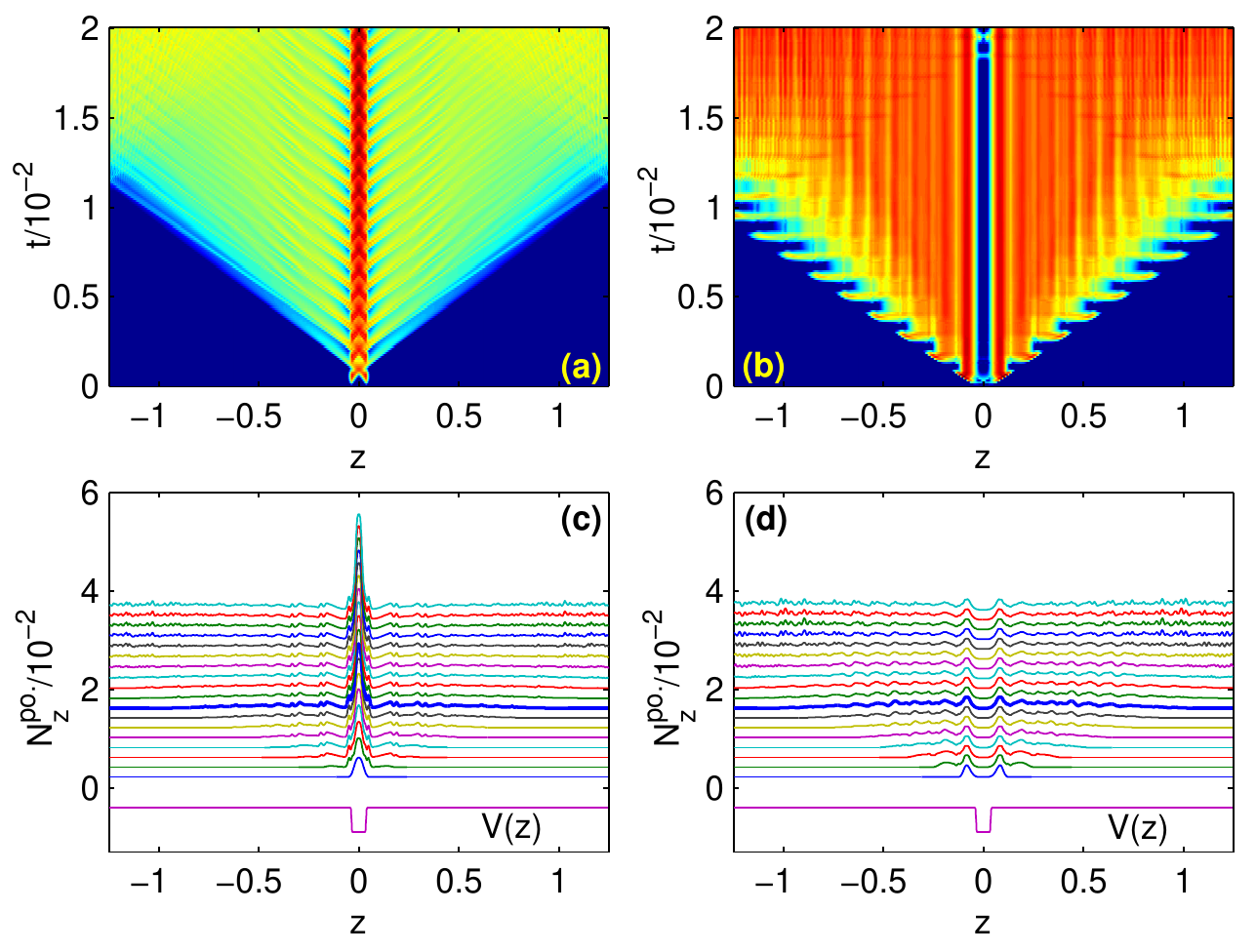}

\protect\caption{For V-oscillating mode, $\omega_{V}=0.3c^{2}$, the three dimensional
diagrams for entire time and the waterfall figures for field free
moments ( the time indicated by triangles in Fig. \ref{fig:N_t} (b)),
for electron spacial density (a, c) and positron spacial density (b,
d). The thicker curve in sub-figure (c, d) mark the last cycle before
the positron arrive the boundary. The well potential $V(z)$ with
$V_{0}=2.53c^{2}$, $W=10\lambda_{C}$, are included on the bottom
for comparison. All other parameters are the same as Fig. \ref{fig:N_t}
(b). \label{fig:imagesc_V}}
\end{figure}

In Fig. \ref{fig:imagesc_W}, for W-oscillating mode, $\omega_{W}=0.3c^{2}$,
we plot the the time evolution of spacial density of electrons and
positrons (sub-figure (a) and (b)). Specially, for the moments when
the field are zero, denoted by the triangles in Fig. \ref{fig:N_t}
(a), these quantities are plotted in the waterfall figures, Fig. \ref{fig:imagesc_W}(c,
d). For V-oscillating mode, $\omega_{V}=0.3c^{2}$, similar diagram
are presented in Fig. \ref{fig:imagesc_V}. For comparison, the well
potential $V(z)$ with wide and depth equal to the upper boundary
of the two modes, $V_{0}=V_{2}=2.53c^{2}$, $W=W_{2}=10\lambda_{C}$,
are included on the bottom. These figures clearly show the process
how are the particles pumped from the well and spread in the numerical
box. 

Since $\omega_{W}=0.3c^{2}$, The period of the width oscillating
is $T_{W}=1.12\times10^{-3}$. Before positrons arrive the boundaries,
the width can finish eight cycles. If we detect the particle population
at the boundary, we can find that positrons arrive the boundary first,
at $t=9.15\times10^{-3}$, in conformity to the estimation $L/\left(2c\right)=9.12\times10^{-3}$.
Electrons arrive the boundary at $t=1.02\times10^{-2}$, about one
period ($T_{W}$ or $T_{V}$ ) later than the positrons. We can see
that the particles reflected by the boundary come back to the interaction
region, and may cause non-ignorable effect, i.e., the non-linearity
of the last three triangles in the dot line in Fig. \ref{fig:N_t}(a),
$\omega_{W}=0.3c^{2}$.

Comparing with the rising and falling bottom of the well, more work
is done by the wall of the well in the case of opening and closing
the well. In the W-oscillating mode, the wavefront of the particles
are more abrupter and regular. In energy space, higher energy modes
are excited, and the spectrum show periodic structure with $0.3c^{2}$
between each peak. In the V-oscillating mode, electrons are lifted
and released naturally. Less work is done and only low momentum mode
are excited, the rate of electrons in the well region ($-5\lambda_{C}<z<5\lambda_{C}$)
is more larger. Also, We can see the absent of interferences both
in or out the well, as discussed in \citep{Dynamics_Bound_States_Su_2005}.

\subsection{time evolution of pumping rate}

\begin{figure}
\includegraphics[scale=0.58]{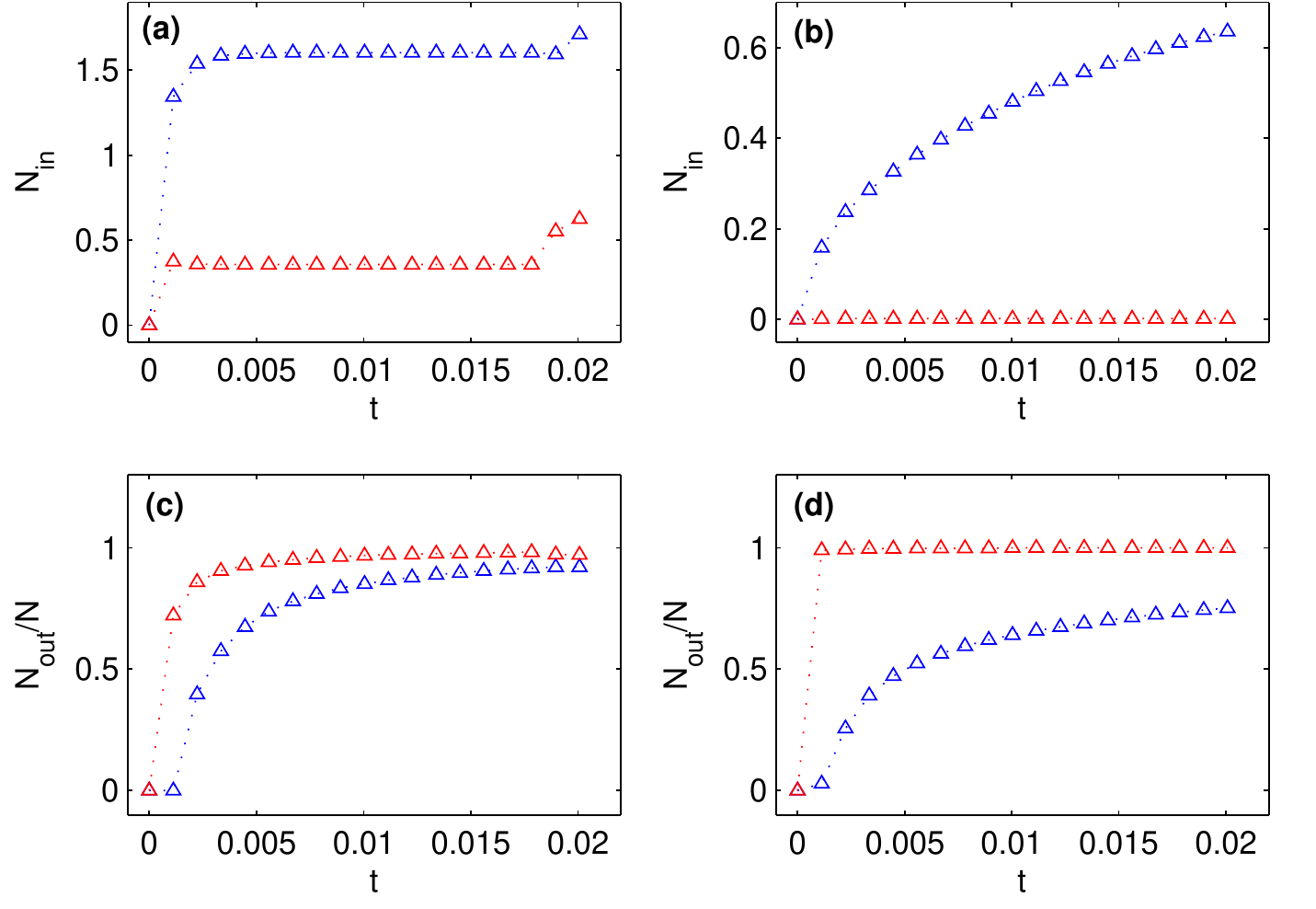}

\protect\caption{For W-oscillating mode ($\omega_{W}=0.3c^{2}$, sub-figure a, c) and
V-oscillating mode ($\omega_{V}=0.3c^{2}$, sub-figure b, d), particles
in the well ($N_{in}$) and the pumping rate $N_{out}/N$ as a function
of time. The triangles denote the time when field is absent and the
dot line link them. The blue triangles denote electron and the red
denote positron. All parameters are the same as Fig. \ref{fig:imagesc_W}
and Fig. \ref{fig:imagesc_V}, respectively.\label{fig:pump_rate}}
\end{figure}

In the V-oscillating mode, it turns out that the electrons is more
inclined to gather in the well region (defined as $-5\lambda_{C}<z<5\lambda_{C}$)
than it in W-oscillating mode . We can integrate the spacial density
$N\left(z\right)$ in this region and get the particle number in the
well, $N_{in}^{el.(po.)}(t)=\intop_{-5\lambda_{C}}^{5\lambda_{C}}N_{z}^{el.(po.)}(t)dz$.
For the pumping process in last section, $N_{in}^{el.(po.)}(t)$ are
graphed in Fig. \ref{fig:pump_rate}(a, b). In W-oscillating mode,
as time increasing, $N_{in}^{el.}$ increase to a constant $1.60$
quickly, while $N_{in}^{po.}$ to a constant $0.36$. But in W-oscillating
mode, $N_{in}^{el.}$ keep increasing while $N_{in}^{po.}$keep zero.
The reason is positrons can be generated in the well region in W-oscillating
mode, while the wall (the electric field) shut the door upon positrons
in V-oscillating mode.

In a pumping process, the pumping rate is vitally important and can
be defined as $\alpha(t)=N_{out}/N$, where $N_{out}=N-N_{in}$, as
shown in Fig. \ref{fig:pump_rate}(c, d). In both modes, at the end
of the first cycle, when $t=T_{W}$ or $T_{V}$, nearly all the electrons
are limited in the well region, while positron are ejected. In V-oscillating
mode, since all the generated positrons are ejected and kept out of
the well, the pump rate directly become $1$. For electron in the
V-oscillating mode, or electron and positron in W-oscillating mode,
in the long time limit, $\alpha(t)$ come to $1$ as $1-\beta/t$,
where $\beta$ depends on the saturation number of particles in the
well and the number of particles can be generated in each cycle.

\subsection{The adiabatic limit}

\begin{figure}
\includegraphics[scale=0.65]{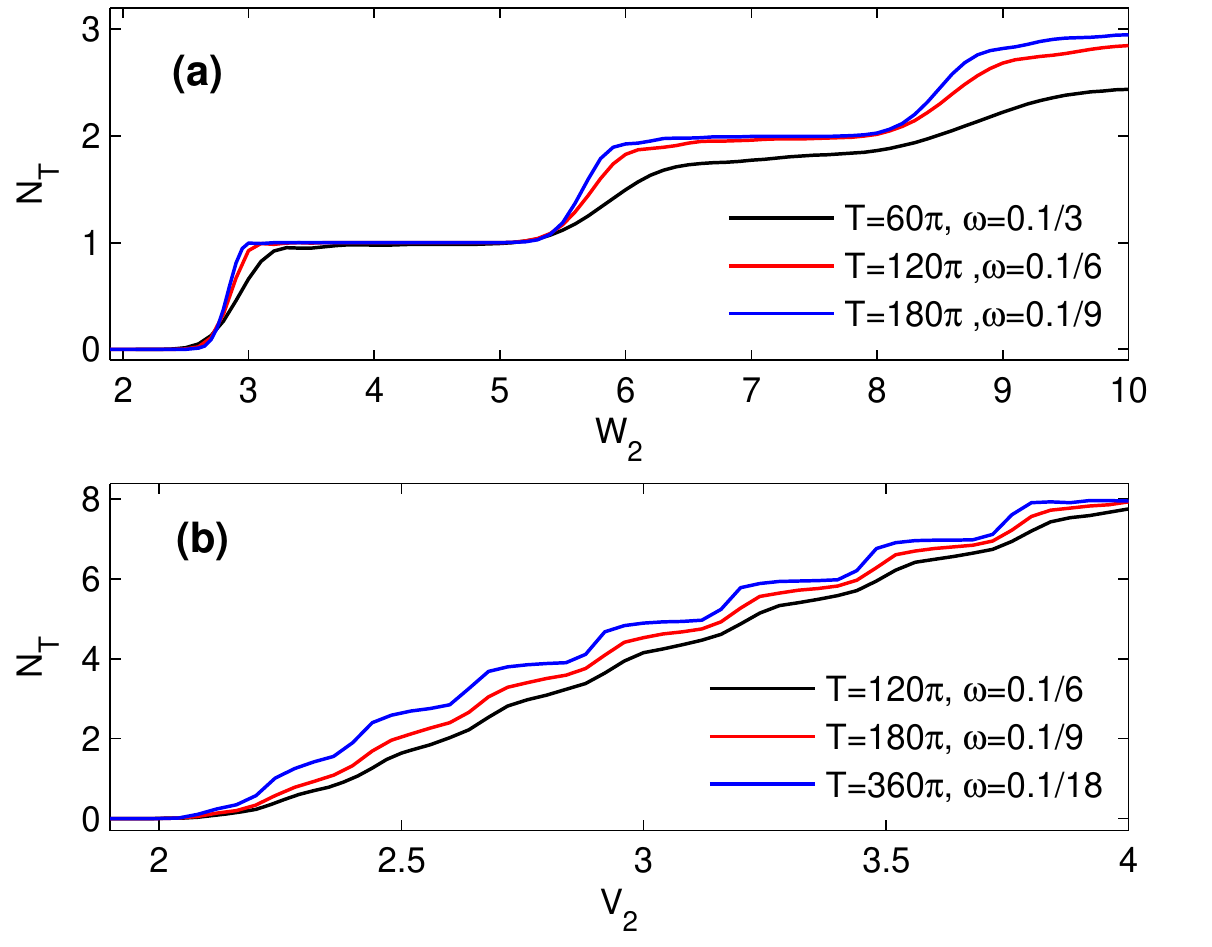}

\protect\caption{The final number of pairs created after one cycle as a function of
the upper boundary of the oscillating width and depth. (a), W-oscillating
mode, $V_{0}=2.53c^{2}$; (b), V-oscillating mode,\textbf{ }$W=10\lambda_{C}$.
The total time $T$ is chosen equal to the oscillating period. $W_{2}$
is in units of $\lambda_{C}$, $T$ is in units of $1/c^{2}$, $\omega$
and $V_{2}$ is in units of $c^{2}$. \label{fig:adiabatic}}
\end{figure}

In Fig. \ref{fig:N_t}, for $\omega_{W}=0.1/6c^{2}$ and $\omega_{V}=0.1/6c^{2}$,
there are $N=2.85$ and $N=1.74$ pairs survived at the end $t=120\pi/c^{2}$.
We have proposed that in low frequency limit, the pairs survived finally
should equal to three, the maximum number of embedded bound states
swept in one cycle of each mode. In Fig. \ref{fig:adiabatic}, ignoring
the reflection, for each frequency, the total time is chosen equal
to the oscillating period for both modes, so that the oscillation
can only finish one cycle. The final number of pairs survived as a
function of the upper boundary of the oscillating width ( $W_{2}$
) and depth ( $V_{2}$ ) are presented. 

In the adiabatic limit, a sub-critical well potential can not trigger
pairs. As the width or depth increasing, the bound states in the gap
dive into the negative continuum successively, the potential become
super-critical. Pairs can be generated and saturated to the number
of embedded bound states. However, as the width and depth decreasing,
bound states depart the negative continuum successively and the generated
pairs can not annihilate because of the non-synchronous ejection.
Finally, the number of pairs survived in this cycle is equal to the
maximum number of bound states embedded. This maximum number is a
function of the upper boundary of the two oscillating cycle ( $W_{2}$
or $V_{2}$ ). 

For a low frequency, the curve which indicate the final number of
pairs vs. $W_{2}$ or $V_{2}$ is like a flight of stairs. As the
frequency become lower, the rising edge of the stairs become more
sharper. As shown in Fig. \ref{fig:adiabatic}, in the limit $\omega_{W},\,\omega_{V}\rightarrow0$,
the the rising edge of the stairs will precisely locate at the points
where the bound states dive into the negative continuum. These points
are $W=2.79,5.51,8.21...$(in units of $\lambda_{C}$), and $V_{0}=2.05,2.19,2.38,2.62,2.87,3.15,3.43,3.73,...$(in
units of $c^{2}$), as illuminated in Fig. \ref{fig:spectrum}. 

The gaps between bound states in the positive and negative continuum
in V-oscillating mode are smaller than that in W-oscillating mode.
To achieve a quasi-adiabatic ( finite $T_{W}$ or $T_{V}$) simulation,
$T_{V}$ should be larger than $T_{W}$ to build a similar stairs. 

Now, if the two quasi-adiabatic oscillating cycle repeate periodically,
we can expect a linear increasing pair number, i.e., for Fig. \ref{fig:adiabatic}(a),
$W_{2}=7\lambda_{C}$, the final pair number will be $2$ times the
number of the cycles.

\section{summary}

In this work, we have constructed a toy model, one-dimensional well
potential with its width and depth oscillating, and studied the electron-positron
creation. Since the bound states diving behavior in the energy spectrum
are similar when sweeping the depth or width, the physical process
are similar in these two modes. We find that the non-synchronous ejection
of particles prevent the particle annihilation, break the Pauli block
effect in a static super-critical well potential, and lead to a high
constant production rate. The width oscillating mode can deliver more
energy to particles and is more efficient in pumping pairs than the
depth oscillating mode. The time evolution of spacial density illustrate
the particles pumping from the well and the spreading of them in the
numerical box. In a quasi-adiabatic case, pair number as a function
of upper boundary of the oscillating, will reveal the diving of the
bound states. This can be expected to detect the energy structure
of a complicated potential. 

In order to reduce the computing cost, we neglect the larger part
of the discrete momentum in the numerical simulation. On the other
hand, with the same number of discrete momentum, the number of spatial
points can be larger to describe the details of the potential. The
simulation in this paper is done on a personal stand-alone computer.
In this algorithm, the time evolution of each negative eigenstate
can be done on a single CPU, hence the computation can be paralleled
easily. Further more, if the second order spatial derivative in the
Hamiltonian are done by finite difference approximations instead of
Fourier transformation\citep{splitoperatormethod}, more lager one-dimensional,
even two dimensional system can be simulated through paralleling the
algorithm on memory shared parallel computers.
\begin{acknowledgments}
This work is supported by National Basic Research Program of China
(973 Program) (Grants No. 2013CBA01502, No. 2011CB921503, and No.
2013CB834100), the National Natural Science Foundation of China (Grants
No. 11374040, No. 11274051, and 11475027).\end{acknowledgments}

\end{document}